\documentclass[prl,amsmath,amssymb,twocolumn,showpacs,floatfix,superscriptaddress]{revtex4}
\usepackage{comment} 
\usepackage{epsfig}
\usepackage[caption=false]{subfig}
\usepackage{float}
\usepackage{graphicx}
\usepackage{mathrsfs,amssymb,amsfonts,amsmath,mathtools}
\usepackage{bbm}
\usepackage{bm}
\usepackage{wasysym}
\usepackage{color}
\usepackage{fancyhdr}

\newcommand{\beq}{\begin{equation}}
\newcommand{\eeq}{\end{equation}}
\newcommand{\nn}{\nonumber}

\newcommand{\anull}{a}

\newcommand{\be}{\beta}

\newcommand{\rmd}{{\rm d}}

\definecolor{mygreen}{rgb}{0,0,0}

\def\be{\begin{equation}}
\def\ee{\end{equation}}
\def\bc{\begin{center}}
\def\ec{\end{center}}
\newcommand{\bea}{\begin{eqnarray}}
\newcommand{\eea}{\end{eqnarray}}

\usepackage[colorlinks,bookmarks=false,citecolor=blue,linkcolor=blue,urlcolor=blue, linktoc=page]{hyperref}
\begin{document}

\title{Nonperturbative phase diagram of interacting disordered Majorana nanowires}
\author{Fran\c cois Cr\'epin}
\affiliation{
Institute for Theoretical Physics and Astrophysics,
University of W\a"urzburg, 97074 W\a"urzburg, German}
\author{G. Zar\'and}
\affiliation{BME-MTA Exotic Quantum Phases 'Lend\"ulet' Group, Institute of Physics, Budapest University
of Technology and Economics, H-1521 Budapest, Hungary}
\author{Pascal Simon}
\affiliation{Laboratoire de Physique des Solides, CNRS UMR-8502,
Universit\'{e}  Paris Sud, 91405 Orsay Cedex, France}

\date{\today}


\begin{abstract}
We  develop a Gaussian variational approach in replica space to investigate the phase diagram of a one-dimensional 
 interacting disordered topological superconducting wire in the strong coupling regime. 
 This method  allows for a non-perturbative treatment in the disorder strength, electron-electron interactions and the superconducting pairing amplitude.
 We find only two stable phases:  a topological superconducting phase,  and a glassy, non-topological localized phase,
 characterized by replica symmetry breaking. 
  \end{abstract}
\pacs{71.10.pm, 74.45.+c, 74.78.Na, 74.81.-g}

\maketitle
{\em Introduction.} Majorana fermions (MF) have recently attracted a lot of attention in condensed matter systems. They are interesting from a fundamental point of view as emergent exotic quasiparticles~\cite{reviews} but also for their potential applications in quantum computing~\cite{rmp}.
In a seminal work, Kitaev~\cite{kitaev} constructed a simple model for a one-dimensional (1D) topological superconductor 
with $p$-wave pairing, hosting Majorana edge states at each end.  Parallel to Kitaev's work, a disordered version of the same 
toy model  has been studied by Motrunich et al, who showed that these  edge states  survive the presence of moderate 
disorder~\cite{Huse}.
Since Kitaev's proposal, many new platforms have been proposed to emulate an effective $p$-wave triplet pairing. Among them, semiconductor wires with spin-orbit and Zeeman interactions in proximity to an $s$-wave superconductor~\cite{lutchyn,oreg} have received considerable attention, as materials are now available experimentally. Transport signatures, which may be explained with Majorana zero modes~\cite{mourik,xu,heiblum}, were reported in recent experiments. 

While  small disorder is harmless~\cite{Huse},  large disorder in a 1D $p$-wave superconductor with 
broken time-reversal and spin SU(2) symmetry (class D) can, however, 
affect the stability of the topological phase and can drive a transition to a non-topological insulating phase~\cite{Huse,disorder}. 
Impurities can also lead to localized, non-trivial, in-gap states 
different from Majorana fermions \cite{disorder-transport},  whose transport signatures may, nonetheless, look much alike~\cite{disorder_exp}. 
Most of the previous studies, however,  ignore the effect of electron-electron interactions, which have been shown to  modify strongly the domain of stability  of the topological phase~\cite{suhas,stoudenmire,rosch,lutchyn_fisher}. Therefore,  one may  wonder whether the topological superconducting (SC) phase  survives when both disorder and electron interactions are taken into account or, eventually,   some new phase emerges. 

The effect of electron-electron interactions on generic disordered  1D $p$-wave superconductors
has been addressed recently  using Abelian bosonization and the perturbative renormalization group (RG) by 
Lobos {\it et al.}~\cite{lobos}. 
They  concluded that there is a quantum phase transition from a topological superconducting phase to a non-topological
localized phase. However, in a large regime of experimental relevance, 
%
including the non-interacting and weakly interacting limits, both  the effective disorder and the proximity induced gap scale 
to  strong coupling at low energies, a regime outside the reach of   perturbative methods.

In this letter, we develop a  non-perturbative low-energy  theory for an interacting disordered $p$-wave SC
using  a  Gaussian variational approach (GVA) in replica space~\cite{parisi}.     
This method  has been shown to capture  the thermodynamic and transport properties of a disordered wire~\cite{GVM} 
and  even more exotic phases such as the Mott glass, an insulating phase with a non-zero optical conductivity 
resulting from the competition between a Mott and an Anderson insulating phase~\cite{orignac}. 
The original approach of M\'ezard and Parisi, however, cannot be applied to the present problem, and needs be 
somewhat modified; while superconductivity  tends to localize the \emph{phase} of the Cooper pairs, 
disorder has a tendency to localize charge carriers in real space and pin their \emph{density},  the conjugate variable of the 
superfluid phase. Therefore, unlike previous theories~\cite{GVM,orignac}, both the superfluid phase and its conjugate momentum 
must be kept in the variational theory and treated at equal footing.  With this modification, the  GVA  is, however, 
 well-suited to capture  the competition between disorder and proximity induced superconductivity.

Within the non-perturbative GVA, we find that only two phases are present: a glassy  Anderson insulating  phase with broken replica symmetry, dominated by disorder, and a replica symmetrical topological SC phase that can support Majorana edge states. We derive the transition line between these two phases. In 
the non-interacting case, we recover using the GVA the transition line  previously obtained by Motrunich {\it et al.} with the real-space renormalization group 
 apporach which assumes  strong disorder \cite{Huse}.

{\em Model.} The low energy physics of a semiconductor nanowire with a strong Rashba spin-orbit interaction $\alpha_R k_F$ ($k_F$ the Fermi momentum), and a Zeeman interaction $E_Z$ in proximity to an $s$-wave superconductor (with induced s-wave pairing $\Delta_S$) can be captured by a model of spinless fermions with  $p$-wave  pairing  \cite{lutchyn,oreg}. Treating  pairing, electron interactions and disorder on  equal footing is a notoriously difficult task. Therefore we follow  Ref.~\cite{suhas} and first diagonalize the nanowire Hamiltonian in the presence of  spin-orbit and Zeeman interactions,  to obtain the  dispersion relation 
$\epsilon_{\pm}(k) = k^2/2m  \pm \sqrt{(\alpha_R k)^2  + (E_Z/2)^2}$ ~\cite{lutchyn,oreg}, with $m$ the electrons'  effective mass (we set 
$\hbar=1$), with  $\pm$ labeling the two bands.
Expanding the singlet SC term in this eigenbasis leads to superconducting order parameters of the
triplet (within the $-$ and $+$ subbands) as well as of the singlet type (mixing $-$ and $+$ subbands).
Majorana edge states require triplet pairing \cite{sato,sau,alicea,lutchyn,oreg},
which is achieved by tuning the chemical potential $\mu$ to lie within the magnetic field gap, 
where only the $\epsilon_-$ subband is occupied, and then turning on a small pairing interaction~\cite{lutchyn,oreg}. 
In this regime,  however, one can describe the  physics of the wire by simply neglecting the empty $+$ band,  
and by  projecting the pairing term, the electron-electron interactions, and the disorder potential to the $\epsilon_-$ subband. 

In the projected theory, the low energy  properties are then  described using standard Abelian bosonization \cite{giamarchi}, with 
the action given as 
$S=S_0+S_\Delta+S_{\rm dis}$ with 
\bea
S_0[\phi,\theta] &=& \int_{0}^{\beta} d\tau \int_{0}^{L} dx \left( -\frac{i}{\pi} \partial_x \theta(x,\tau) \partial_\tau \phi(x,\tau) \right. \\ 
&+& \frac{v}{2\pi} \left[ \left. K (\partial_x \theta (x,\tau))^2 + \frac{1}{K}(\partial_x \phi (x,\tau))^2 \right]  \right)\;,\label{eq:s0}\nn\\
S_\Delta&=&  - \frac{2 \Delta}{\pi \anull} \int dx d\tau \; \cos[2\theta(x,\tau)], \label{eq:sp}\\
S_{\rm dis}&=&- \frac{1}{2\pi \anull} \int dx d\tau \; \left[ \xi(x)e^{2i\phi(x,\tau)}+{\rm H.c.}\right]\;.\label{eq:sdis}
\eea
 The first term  $S_0$ describes the physics of the interacting electron fluid in the $\epsilon_-$ band, 
 with   collective plasmon excitations of velocity $v$.  
 The bosonic 'displacement' field $\phi$ (more precisely $\partial_x\phi$) 
 represents  charge density fluctuations, while the phase field $\theta$ is conjugate to it, 
 as expressed by the commutation relation $[\phi(x),\theta(x')]=-i\frac{\pi}{2}{\rm sign}(x-x')$. 
The Luttinger parameter $K$ encodes electron-electron interactions with $K<1$ (resp. $K>1$) for repulsive (resp. attractive) 
interactions. 
The second term of the action, $S_\Delta$ describes triplet pairing. Here $a$ denotes a short length cut-off, and the effective 
pairing interaction, $\Delta\approx \Delta_S (\alpha_R k_F)/E_Z$,
tries to pin the superfluid phase to the minima $\theta = n \pi $.
Finally, the last term $S_{\rm dis}$ describes backscattering on  a Gaussian quenched disorder,
 $\xi(x)$, satisfying  $\overline{\xi(x) \xi^*(x')}= D\, \delta(x-x')$. In $S_{\rm dis}$
we dropped  forward scattering terms,  since they can be gauged away and do not contribute to localization~\cite{giamarchi}.

{\it The Gaussian variational method.}
To make progress, we use the replica trick~\cite{parisi}: we introduce $n$ copies of the 
fields $(\phi,\theta)\to (\phi^a,\theta^a)$ with $a\in [1,n]$, average over the Gaussian disorder,
 and finally take the limit $n\to 0$. The replicated action thus obtained reads as
\bea
S&= &\sum_{a=1}^n  S_0[\phi^a,\theta^a]  - \frac{2 \Delta}{\pi \anull} \int \rmd x\, \rmd \tau \; \cos[2\theta^a(x,\tau)] 
\label{eq:S}
\\
&- &\frac{D}{(2\pi \anull)^2} \sum_{a,b =1}^n \int \rmd x \,\rmd\tau \,\rmd\tau' \cos 2(\phi^a(x,\tau) - \phi^b(x,\tau')). \nn
\eea
As revealed by a simple perturbative RG analysis~\cite{lobos},  the RG eigenvalues  of the pairing and disorder 
terms are simply $2-K^{-1}$ and  $3-2K$, respectively, and are  both relevant for $1/2\leq K\leq 3/2$, where
 the system flows to  strong coupling.  Lobos et al.~\cite{lobos}  conjectured therefore two phases, separated by a critical line:
a topological SC phase dominated by  $\Delta$, and a non-topological Anderson-localized phase, driven by  $D$. 

To establish  the  Gaussian variational approach (GVA), we first rewrite the first term 
of the  action Eq.~\eqref{eq:S}  
in Fourier space $Q = (i\omega_n,q)$ as 
$S_0[\Psi] = \frac{1}{2 \beta L}\sum_Q  \Psi^{\dagger}(Q) G_0^{-1}(Q) \Psi(Q)$,
with $\Psi^T = (\theta^1,\ldots,\theta^n,\phi^1,\ldots,\phi^n)$  and
\beq
G^{-1}_0(Q) = 
\begin{pmatrix}
[G^{-1}_0(Q)]_{11}  & [G^{-1}_0(Q)]_{12}  \\
[G^{-1}_0(Q)]_{21}  & [G^{-1}_0(Q)]_{22} 
\end{pmatrix}. \label{Eq:G0_new}
\eeq
The $n\times n$ blocks $[G^{-1}_0(Q)]_{ij}$, $i,j=1,2$, are given by $[G^{-1}_0(Q)]_{11}= vK \,q^2/\pi$, $[G^{-1}_0(Q)]_{12} = [G^{-1}_0(Q)]_{21} = i q \omega_n /\pi$ and $[G^{-1}_0(Q)]_{22}= (v/K) \,q^2/\pi$ .  The basic idea of the GVA is to approximate the action $S$ of Eq.~\eqref{eq:S} by the best Gaussian action in replica space, 
 $S\to S_G[\Psi]= \frac{1}{2 \beta L}\sum_Q \Psi^\dagger(Q) G^{-1}(Q) \Psi(Q)$, with 
$G^{-1}$  a $2n\times 2n$ matrix in replica space. Using the well-known inequality, 
$F \le F_{\rm var}[G] \equiv F_G + k_B T \,\langle
S-S_G\rangle_G$, and minimizing the variational free energy
$F_{\rm var}$ with respect to $G$ we obtain self-consistency equations for $G(Q)$ and, at the same time, an estimate for 
the free energy $F$.
Details of the calculations are presented in the supplemental material \cite{supmat}. 
As was shown in Ref.~\cite{GVM}, the  localized phase of
fermions in a disordered potential is described by a replica symmetry breaking (RSB) solution. Therefore, we expect 
the emergence of a replica symmetry broken solution in the disorder-dominated phase. 
The superconductivity  dominated phase is, however, expected to be replica symmetric (RS).
These expectations are indeed confirmed by the detailed analysis presented below.
The phase transition between these two phases is, however, found to be of first order, implying that
close to the phase boundary both phases are locally stable. 
Our strategy is therefore to find both the RS and  RSB solutions, and  compare their free energy 
to determine the stable phase of minimal  free energy. 

Physical quantities as well as the self-consistency equations involve the 
so-called  connected propagator, $G^c_{ij}\equiv  \sum\limits_{b}G^{ab}_{ij}$, 
and its inverse, $(G^{-1})^c_{ij} =  (G^c)^{-1}_{ij}  \equiv  \sum\limits_{b}(G^{-1})^{ab}_{ij}$.
The propagator $G^c(Q)$ can be shown to be simply  the disorder-averaged correlation function of 
the fields $\theta $ and $\phi$ and, within the GVA, we can express it as   
\beq
(G^{-1})^c(Q) = \frac{1}{\pi}
\begin{pmatrix}
vK q^2  + m & i q \omega_n \\
i q \omega_n & v \,q^2/K + \Sigma(\omega_n)
\end{pmatrix} \;.
\eeq
A finite value of the renormalized gap, $m$, in this equation sustains superconducting order, while 
RSB, i.e., a glassy phase manifests itself through the  non-analytical structure of  
the connected self-energy $\Sigma(\omega_n)$.

{\em Superconducting phase: {\rm RS} self-consistency equations.} 
The replica-symmetrical ansatz always leads to a regular self-energy, 
 $\Sigma(\omega_n) = I(\omega_n)$, and describes a  superfluid (or normal) phase. 
 Introducing the variables $ \hat{I} \equiv (\pi K /v)  \,I$ and $\hat{m} \equiv \pi \,m /(vK) $,
 the self-consistency equations for the renormalized superconducting gap 
reduce to  
\bea
m &= & \frac{8 \Delta}{\pi \anull} 
e^{ -2 [G^c]_{11}(x=0,\tau=0)} \;, 
\label{eq:m}
\\
{[G]}^c_{11} (x=0,\tau=0) &= &\frac{1}{\beta L} \frac{\pi}{vK} \sum_Q \frac{q^2 + \hat{I}(\omega_n)}{D(Q,\hat{I})}\;,
\label{eq:Gc11}
\eea
with the denominator defined as 
\beq
D(Q,\hat{I})=(q^2 + \hat{m})(q^2 + \hat{I}(\omega_n))+ q^2\omega_n^2/v^2\;.
\label{eq:D}
\eeq
For $1/2\le K$, these self-consistency equations always yield  a solution with a finite mass $m$,  
 corresponding to the topological superconducting phase. Surprisingly, we have not found  any indication 
 for a breakdown  of this RS solution in the function $I(\omega)$, and concluded that this solution 
 appears to be locally stable against replica symmetry breaking. 

{\em Localized phase: RSB self-consistency equations.}
The self-consistency equations also admit a stable,  one-step replica symmetry-breaking solution (1RSB), 
similar to the ones found in Refs.~\cite{GVM} and \cite{crepin}. This solution can be interpreted as a
glassy, interacting localized phase.  In this phase, the replica structure of the Green's functions and thus the 
self-consistency equations are more involved~\cite{supmat}. They can be constructed by using
the  Parisi-parametrization of the replica-matrices, $G^{ab}_{ij}$. 
In this localized phase, the self-energy is found to  develop a singular structure, 
 \[
 \Sigma(\omega_n) = I(\omega_n) + \Sigma(1-\delta_{n,0}),
 \]
with the RSB appearing through the non-zero value of $\Sigma>0$. The last term in this expression 
generates a length scale, $\xi\equiv {\hat\Sigma}^{-1/2}=({\Sigma\,\pi K/v})^{-1/2}$, and a corresponding energy scale, 
$\omega_\xi\equiv v/\xi$, which can be  identified as the \emph{localization length} and a corresponding \emph{pseudogap}, respectively~\cite{GVM}.

We analyzed the self-consistency equations within the simplest  approximation, $I(\omega) \to  0$~\cite{GVM}. 
In this limit, they reduce to a set of two coupled integral equations for $m$ and $\Sigma$. 
Surprisingly,  however, we found no solutions with $m \neq 0$ and  $\Sigma \ne 0$. This result has the important consequence, that 
the two 'order parameters', $\Sigma$ and $m$ are  {\em mutually exclusive} within the GVM, i.e.,  
no exotic phase  analogous to a Mott glass phase~\cite{orignac} or the Bose-Fermi glass phase~\cite{crepin} emerges, in agreement with the conjecture of Ref.~\cite{lobos}.


\begin{figure}[t] 
\centering
\includegraphics[width=8.6cm,clip]{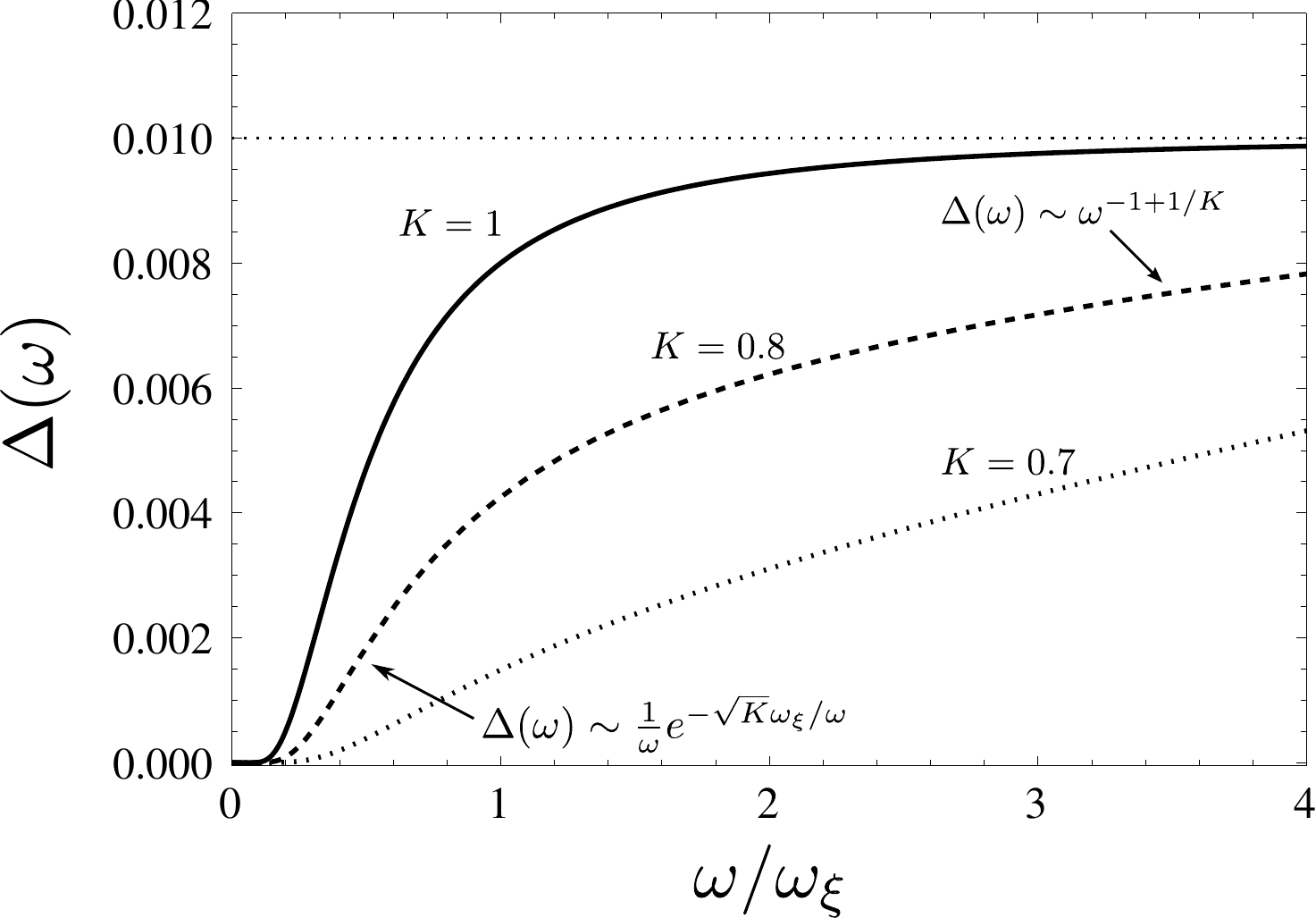}
\caption{Effective gap ${\Delta}(\omega)$ as a function of $\omega/\omega_\xi$ 
for  $\Delta_0\; a/v = 0.01$,  $\hat \Sigma \,a^2 = 0.01$ and different values of the Luttinger parameter, $K$.
A pseudogap feature emerges below the scale $\omega_\xi$.
  }
\label{Fig:flow1}
\end{figure}

To confirm this result, we performed a RG analysis around the 1RSB fixed point by adding 
a  pairing potential to the Gaussian 1RSB action,  $S_{1\rm RSB}$
\beq
S \equiv  S_{1\rm RSB} + \sum_{a=1}^n   \tilde{\Delta} \frac{v}{\anull^2}\int  {\rm d}x \, {\rm d}\tau \cos[2\theta^a(x,\tau)],
\nonumber
\eeq
and constructing the scaling equation for  
 the dimensionless pairing, $\tilde{\Delta} = \Delta \anull/v $. The anomalous dimension of $\tilde{\Delta}$ is 
 now scale dependent and, correspondingly,    the effective  pairing potential at energy $\omega$, 
 ${\Delta}(\omega) = \omega \tilde{\Delta}(\omega)$ is found to obey the following scaling  equation:
\beq
\frac{{\rm d} \ln {\Delta}(\omega)  } {{\rm d}\ln(\Lambda/\omega)} = 1 - \frac{\sqrt{\omega_\xi^2+ \omega^2} }{K\;\omega}\;,
\label{eq:DeltaScaling}
\eeq
with $\Lambda\approx  v/ a$ a high energy cut-off. 
At high frequencies $ {\Delta}(\omega)$  therefore  behaves as 
${\Delta}(\omega)\sim \omega^{-(1-1/K)}$, while at small frequencies, 
$\omega < \omega_\xi$, it scales   exponentially  to zero, ${\Delta}(\omega)  \sim \frac1{\omega} \,e^{- {K}^{-1/2}\, \omega_\xi/\omega }$,
consistently with the value of  $m=0$ we get from the variational calculation. Therefore, even though $\Delta$ may appear to be a relevant 
perturbation at high energies, a finite localization length turns it to be irrelevant at small frequencies, and drives it effectively to zero (see
Fig.~\ref{Fig:flow1}).

We thus conclude from the GVM analysis  that --   in concordance with  the perturbative RG approach of Ref.~\cite{lobos} -- 
only two mutually exclusive phases emerge  in the strong coupling regime for $1/2\leq K\leq 3/2$: a glassy phase with a finite localization length, 
and a topological superconducting phase, separated from the former  phase by a first order transition.

\begin{figure}
        {\includegraphics[width=0.47\textwidth]{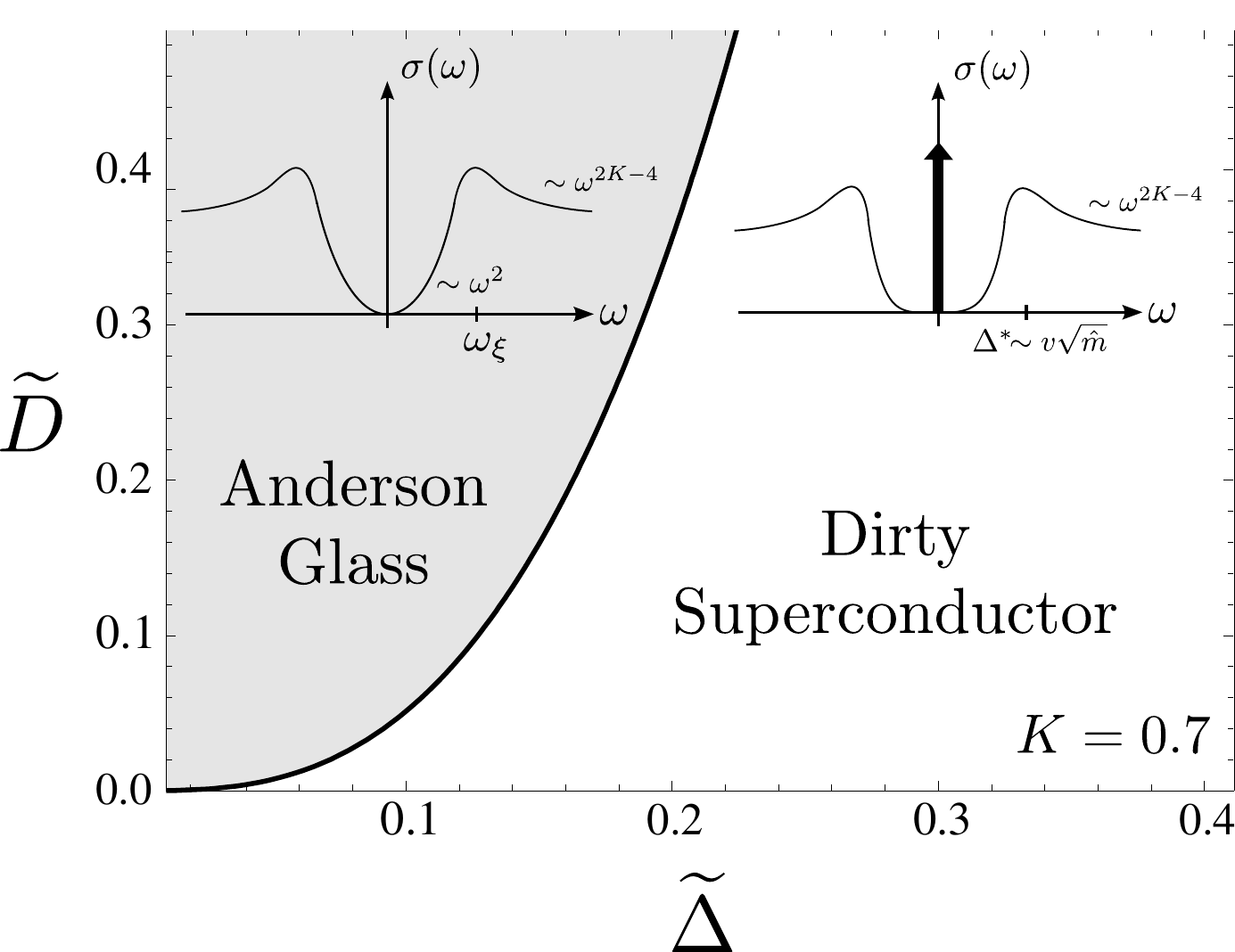}}
		\caption{Phase diagram as a function of  the dimensionless disorder parameter, $\widetilde D$, and  the dimensionless pairing potential, 
		$\widetilde \Delta$, for a repulsive interaction ($K=0.7$), as obtained from the free-energy calculation. A repulsive interaction stabilizes 
		the glassy phase. The two phases are separated by a first-oder line. The insets sketch the optical conductivity $\sigma(\omega)$ 
		in both phases.	
		}\label{fig:phase}
\end{figure}

{\em Phase diagram.}
To determine the phase  boundary, we  computed  the  variational free energy in both phases.
The calculations are detailed in the Supplemental material \cite{supmat}.
In the superconducting phase, we obtain the simple closed expression
for the free energy density
\beq \Delta f_{\rm SC} =f-f_{SC}= \frac{v \hat{m}}{8\pi}(1-2K),
\eeq 
 while  in the glass phase we find 
\beq
\Delta f_{\rm AG} = \frac{v \hat{\Sigma}}{8\pi}(1-3/K)\; . \eeq
Defining the dimensionless disorder as $ \widetilde{D}\equiv D a /v^2 $, and using 
$\hat{m}= a^{-2} (8 \widetilde{\Delta}/K)^{\frac{K}{2 K-1}}$ and $\hat{\Sigma}=a^{-2} (2 \widetilde{D} K^2)^{\frac{2}{3 -2K}}$
obtained from the self-consistency equations (valid for $\hat{m} \ll a^{-2}$ and $\hat{\Sigma} \ll a^{-2}$), 
and finally comparing 
$\Delta f_{\rm AG}$ and $\Delta f_{\rm SC}$,  we obtain the phase diagram 
 in Fig.~\ref{fig:phase}. The  phase boundary is given by 
\beq\label{eq:boundary}
\widetilde{D} = \frac{1}{2 K^2}\left( \frac{8 \widetilde{\Delta}}{K} \frac{1-2K}{1-3/K} \right)^{\frac{3-2K}{2-1/K}}\;,
\eeq
and it changes concavity at $K=1$.  For repulsive interactions ($K<1$), in particular, interactions tend to stabilize 
the glassy phase, while for attractive interactions they support the SC phase. 
In the non-interacting case, corresponding to $K=1$, Eq. (\ref{eq:boundary}) simplifies to $\frac{\widetilde{\Delta}}{\widetilde{D}}=\frac{1}{2}$.
This result is exactly the one derived by Motrunich {\it et al.} using a real space renormalization group approach which is suitable in the strong disorder regime. The fact that we recover this result by comparing the variational free energy in both phases from our GVA is highly non-trivial
and indicates that the GVA is able to capture the right physics in the strong coupling regime for this system.
We presented the phase diagram in Fig. \ref{fig:phase} for $K=0.7$,   corresponding to repulsive interactions. 



{\em Observables.} 
While the topological superconducting phase supports Majorana fermion edge excitations, 
the localized glassy phase has no single-particle gap and therefore cannot support such protected Majorana edge states. 
 Many proposals have been made to detect signatures of these Majorana  edge states by means of transport measurements 
 which are sensitive to edge excitations (see \cite{reviews}). However, disorder may  also lead to localized
in-gap states dfferent from Majorana fermions, whose transport signatures may look
much alike Majorana edge excitations \cite{disorder}.
In order to distinguish between the two phases depicted in Fig.  \ref{fig:phase}, one may also rely on bulk observables such as the finite frequency optical conductivity $\sigma(\omega)$. For the 1D Anderson glassy phase, such quantity has been computed in Ref.\cite{GVM}
within the GVA and shown to behave as $\sigma(\omega)\sim \omega^2$ at low frequency. A power counting argument shows instead that
$\sigma(\omega)\sim \omega^{2K-4}$ at large frequency. We therefore expect a maximum of $\sigma(\omega)$ at the scale $\omega_{\xi}$ corresponding to the pseudo gap energy scale. 

In contrast, the topological (dirty) superconducting phase is characterized by a zero-frequency peak in $\sigma(\omega)$ signaling that  superconductivity sets in. At large frequency, the optical conductivity is insensitive to superconducting correlations and therefore $\sigma(\omega)$ behaves similarly in both phases. Though we have not been able to obtain a closed form for the low-energy behavior of $\sigma(\omega)$ in the dirty superconducting phase within the GVA,
we expect it to exhibit  
a maximum around the energy scale $\widetilde\Delta$. Based on these considerations, we have sketched the expected behavior of the optical conductivity in both phases in   Fig.  \ref{fig:phase}.

{\em Conclusions.} We developed  a Gaussian variational theory in replica space
for  a 1D topological superconducting wire in presence of  electron-electron interaction and
disorder. This approach allowed us to capture the competition between pairing and disorder 
in a non-perturbative way. Two stable phases were found: a topological superconducting phase, and a glassy, non-topological phase 
with localized carriers, in concordance with the  phase diagram conjectured  by Lobos et al. \cite{lobos}. 
In the glassy phase, p-wave  superconductivity was shown to be irrelevant at very small energies. 
The phase boundary between   these  two phases has been determined analytically.

{\em Acknowledgments.}
This research has been supported by  the French ANR DYMESYS (ANR 2011-IS04-001-01), 
 the Hungarian Research Fund OTKA and the NF\"U
 under grant Nos. K105149 and CNK80991, respectively.






\onecolumngrid
\newpage
\begin{center}
{\Large \bf Supplementary material for ``Majorana fermions in interacting disordered topological superconducting wires''}
\end{center}

\setcounter{figure}{0}
\renewcommand{\thefigure}{S\arabic{figure}}

\setcounter{equation}{0}
\renewcommand{\theequation}{S\arabic{equation}}


\section{Derivation of the self-consistent variational equations}
We summarize here the general formalism of the GVM which led to the set of self-consistent equations both in the superconducting phase (which is replica symmetric)
and in the localized phase (described by a 1-step RSB scheme).\\

In the inverse Green's function $G^{-1}$, we separate the free, Luttinger liquid, part and a self-energy contribution coming from both the pairing and the disorder potentials, such that $G^{-1} = G_0^{-1} - \sigma$. Here, $G^{-1}$, $G_0^{-1}$ and $\sigma$ are all $2n \times 2n$ matrices, with $n$ the number of replicas. We introduce a double index notation by defining 
\beq
[G(Q)]_{ij}^{ab} = \langle \Psi_i^a(Q)  \Psi_j^b(-Q)\rangle_G\;,
\eeq
where $i,j = 1,2$ and $a,b$ run from $1$ to $n$, with $\Psi_1^a = \theta^a$ and $\Psi_2^a = \phi^a$. We also use the short-hand notation $Q=(i\omega_n,q)$. $G^{-1}$, $G_0^{-1}$ and $\sigma$ are therefore $2\times 2$ block-matrices in the space of $\theta$ and $\phi$, with $n\times n$ blocks (see, e.g. Eq.~\eqref{Eq:G0_new}). The variational free energy has the form

\beq
F_{var} =  - \frac{1}{2\beta}\sum_{Q} \ln \textrm{det} G(Q) +\frac{1}{2}\sum_{i,j,a}\sum_{Q }\left[G_0^{-1}\right]_{ij}^{aa}(Q)G_{ij}^{aa}(Q) + \frac{1}{2} \sum_{a,b}  L \int d\tau \left[ V[B^{ab}(\tau)] - \frac{2\Delta}{\pi \anull} e^{-2G_{11}^{aa}(x=0,\tau=0)} \right]\;, \label{Eq:var_F}
\eeq
with 
\beq
B^{ab}(\tau) = \langle \left[ \phi^a(x,\tau) - \phi^b(x,0) \right]^2 \rangle_G\;,
\eeq
and $V(x) = -2\frac{D}{(2\pi \anull)^2}e^{-2x}$. Differentiation of \eqref{Eq:var_F} with respect to $G$ yields the saddle-point equations. In particular we find that the self-energy matrix $\sigma$ is diagonal in field space, and has the following form in replica space,

\begin{align}
\sigma_{22}^{ab}(Q) &= \left\{
\begin{array}{ll}
2\int_0^\beta d\tau \left(1-\cos[\omega_n\tau]\right) V'(B^{aa}(\tau))+ 2\int_0^\beta d\tau \sum_{b\neq a} V'[B^{ab}]\;, \quad (a=b) \\ \\
-2\beta \delta_{n,0} V'(B^{ab}) \quad  (a\neq b)\;,
\end{array}
\right.  
\\ \nn \\
\sigma_{11}^{aa}(Q) &= -\frac{8 \Delta}{\pi \anull} e^{ -2 [G^c]_{11}(x=0,\tau=0)} \delta_{ab} \;, 
\end{align}
Note that, in the case of static disorder,
off-diagonal quantities, as $B^{ab}$ with $a\neq b$, do
not depend on time~\cite{GVM}. This is because
off-diagonal elements describe correlations between replicas locked 
to different minima, but experiencing the same disorder.  
The experienced random potential being static, these correlations 
are also time-independent. \\

The next step is to take the limit $n \rightarrow 0$. We follow
Parisi's parameterization of $0\times 0$ matrices \cite{parisi}.
If $A$ is a matrix in replica space, taking $n$ to $0$ it can be parameterized by a couple
$(\tilde{a},a(u))$, with $\tilde{a}$ corresponding to the 
replica-diagonal elements and $a(u)$ a function of $u\in [0,1]$, 
parameterizing the off-diagonal elements. To proceed, it is necessary to make an assumption on the form of the self-energy. We actually already know various limits. In the absence of a pairing potential, the localized phase is described by a so-called 1-step replica symmetry breaking (1RSB) solution, for which there exists a value $0<u_c<1$ such that $\sigma_{11}(u<u_c)=0$ and
$\sigma_{11}(u>u_c) = \sigma$, or equivalently $B(u<u_c)=\infty$ and
$B(u>u_c)=B$. In the case where a pairing potential is indeed present we whish to extend this solution by including a mass in the $\theta$ sector. We therefore define $m = - \widetilde{\sigma}_{11}$, which is independent of $Q$. The self-consistent equations take on a simple form, provided we introduce a few auxillary functions that appear naturally in the expression of the so-called connected inverse Green's function
\beq
[G^{-1}]_{ij}^c \equiv \sum_{b \neq a} [G^{-1}]_{ij}^{ab}\; =  \widetilde{G^{-1}_{ij}} - \int_0^1 du \; G^{-1}_{ij}(u)\;.
\eeq
In the case of a 1RSB solution we have
\begin{align}
(G^{-1})^c_{11} &=  (G_0^{-1})_{11} + m \;, \\
(G^{-1})^c_{22} &=  (G_0^{-1})_{22} + I(\omega_n) + \Sigma(1 - \delta_{n,0})\;, \\ 
(G^{-1})^c_{12} &= (G^{-1})^c_{21} = (G_0^{-1})_{12}  \;.
\end{align}
with
\begin{align}
&m = \frac{8 \Delta}{\pi \anull} e^{ -2 \widetilde{G_{11}}(x=0,\tau=0)} \;, \label{eq:m} \\
&I(\omega_n) =  \frac{ 2 D}{\pi^2 \anull^2}   \int_0^\beta d\tau (1-\cos (\omega_n \tau) ) \left( e^{- 2 \widetilde{B}(\tau)} - e^{- 2 B} \right)\;. 
\end{align}
The pseudo mass $\Sigma$ is given by $\Sigma = 2\beta u_c V'[B]$. An extra equation is needed in order to fix the break-point $u_c$. It can be obtained by studying the stability of the variational solution~\cite{GVM,crepin}. It turns out that $\phi$ and $\theta$ decouple in the stability analysis and the break-point is the same as in the case of the Bose glass. The final equation for $\Sigma$ reads:
\beq
(\Sigma)^{3/2} = \frac{ 2 D}{\pi^2 \anull^2}  \sqrt{\frac{\pi K}{v}} e^{- 2 B}\;.
\eeq
In order to close the system we also need the following expressions
\beq
 \widetilde{G_{11}}(x=0,\tau=0) = \frac{1}{\beta L} \frac{\pi}{vK} \sum_Q \frac{q^2 + \hat{I}(\omega_n) + \hat{\Sigma} }{(q^2 + \hat{m})(q^2 + \hat{I}(\omega_n) + \hat{\Sigma})+ q^2\omega_n^2/v^2}\;,
\eeq

\beq
 \widetilde{B}(\tau) = \frac{2}{\beta L} \frac{\pi K}{v} \sum_Q (1-\cos (\omega_n \tau) )\frac{q^2 +  \hat{m} }{(q^2 + \hat{m})(q^2 + \hat{I}(\omega_n) + \hat{\Sigma})+ q^2\omega_n^2/v^2}\;,
\eeq
and
\beq
B = \frac{2}{\beta L} \frac{\pi K}{v} \sum_Q \frac{q^2 +  \hat{m} }{(q^2 + \hat{m})(q^2 + \hat{I}(\omega_n) + \hat{\Sigma})+ q^2\omega_n^2/v^2}\;,
\eeq
where we have introduced $\hat{\Sigma} = (\pi K/v)\Sigma$, $ \hat{I} = (\pi K /v) I$ and $\hat{m} = \pi /(vK) m$. We found no solution of these equations for which both $m$ and $\Sigma$ are non zero. \\

In the case $\Sigma = 0$, the solution becomes replica-symmetric (RS) and the equations simplify to
\begin{align}
(G^{-1})^c_{11} &=  (G_0^{-1})_{11} + m \;, \\
(G^{-1})^c_{22} &=  (G_0^{-1})_{22} + I(\omega_n)\;, \\ 
(G^{-1})^c_{12} &= (G^{-1})^c_{21} = (G_0^{-1})_{12}  \;,
\end{align}
with
\beq
  m = \frac{8 \Delta}{\pi \anull} e^{ -2 [G^c]_{11}(x=0,\tau=0)} \;, 
\eeq
and
\beq
I(\omega_n) =  \frac{ 2 D}{\pi^2 \anull^2}  e^{ -2 [G^c]_{22}(x=0,\tau=0)} \int_0^\beta d\tau (1-\cos (\omega_n \tau) ) \left( e^{4 [G^c]_{22}(x=0,\tau)} -1 \right)\;,
\eeq
as well as 
\beq
[G^c]_{11}(x=0,\tau=0) = \frac{1}{\beta L} \frac{\pi}{vK} \sum_Q \frac{q^2 + \hat{I}(\omega_n)}{(q^2 + \hat{m})(q^2 + \hat{I}(\omega_n))+ q^2\omega_n^2/v^2}\;,
\eeq
and
\beq
[G^c]_{22}(x=0,\tau) = \frac{1}{\beta L} \frac{\pi K}{v} \sum_Q e^{i\omega_n \tau} \frac{q^2 + \hat{m}}{(q^2 + \hat{m})(q^2 + \hat{I}(\omega_n))+ q^2\omega_n^2/v^2}\;.
\eeq
$G^c$ is by definition the inverse of $(G^{-1})^c$.

\section{Calculation of the free energy}
\subsection{Superconducting phase}

The variational free energy is given by:
\beq
F = F_G + \frac{1}{\beta}\langle S - S_G \rangle_G \;.
\eeq
For convenience, we will compute the free energy per unit volume and replica, that is, $f = F/(nL)$. Using obvious notations we decompose the action as $S = S_0 + S_\Delta + S_D$ (see Eq.~\eqref{eq:S}). With our variational solution in the RS phase, $\langle S_D \rangle_G =0$ . We are left with three terms to compute:

\begin{align}
f_1 &= F_G/(nL) \;, \\
f_2 &= \frac{1}{n \beta L}\langle S_0 - S_G \rangle_G \;, \\
f_3 &= \frac{1}{n \beta L}\langle S_\Delta \rangle_G \;. \\
\end{align}
The most straight-forward term is $f_3$. Indeed we have
\beq
f_3 =- n \frac{1}{n\beta L} L\beta  \frac{2 \Delta}{\pi \anull}  \langle e^{2i\theta^a(x,\tau)}\rangle_G = -\frac{m}{4}\;,
\eeq
where we have used Eq.~\eqref{eq:m}. Not so much complicated is $f_2$:
\bea
f_2 =&& - \frac{1}{2 L} \sum_{q,\omega_n}  \frac{\hat{m}}{\frac{\omega_n^2}{v^2}+q^2+\hat{m}} \nonumber\\
=&& -\frac{v \hat{m}}{2L} \sum_{q > 0} \frac{ \coth (\beta v \sqrt{q^2+\hat{m}}/2)}{\sqrt{q^2+\hat{m}}} \;.
\eea
Finally, we are left with the computation of $f_1$. To avoid ambiguities in the definition of the measure in the path integral, we substract $n F_0$ to the free energy, with $F_0$ the free energy of a Luttinger liquid. We are left to compute $\Delta f_1 = f_1 -f_0$. We have:
\begin{align}
\Delta f_1 &= - \frac{1}{\beta L} \ln \left[ \textrm{Tr} e^{-S_G} \right] + \frac{1}{\beta L} \ln \left[ \textrm{Tr} e^{-S_0} \right]  \;, \\
&= - \frac{1}{2\beta L} \sum_{q,\omega_n} \left[ \ln \det G(q,\omega_n) - \ln \det G_0(q,\omega_n)\right]\; \\
&= - \frac{2}{\beta L} \sum_{q >0} \ln \left[ \frac{\sinh(\beta  v q/2)}{\sinh(\beta  v \sqrt{q^2 + \hat{m}}/2)} \right] \;.
\end{align}
Finally, we have for, $\Delta f = f - f_0$, 
\beq
\Delta f =  - \frac{2}{\beta L} \sum_{q >0} \ln \left[ \frac{\sinh(\beta  v q/2)}{\sinh(\beta  v \sqrt{q^2 + \hat{m}}/2)} \right] -\frac{v \hat{m}}{2L} \sum_{q > 0} \frac{ \coth (\beta v \sqrt{q^2+\hat{m}}/2)}{\sqrt{q^2+\hat{m}}} - \frac{\hat{m}}{4} \frac{vK}{\pi}\;.
\eeq
In the limit $\beta \to \infty, L \to \infty$, the two first terms conspire to give a finite integral over $q$, and we are left with a particularly simple expression:
\beq
\Delta f =  \frac{v\hat{m}}{8\pi}(1-2K)\;.
\eeq
$\Delta f <0$ for $K>1/2$, which is consistent with the RG.

\subsection{Localized phase}
We proceed with a similar calculation in the 1RSB phase. The three pieces to compute are now
\begin{align}
\Delta f_1 &= F_G/(nL) - f_0 \;, \\
f_2 &= \frac{1}{n \beta L}\langle S_0 - S_G \rangle_G \;, \\
f_3 &= \frac{1}{n \beta L}\langle S_D \rangle_G \;. 
\end{align}
Again, with start with $f_3$ and find
\bea
f_3 =&& - \frac{D}{\beta L (2\pi \anull)^2}  L \beta \int_0^\beta d\tau  \;  e^{-2\widetilde{B}(\tau)} + (1-u_c) \frac{D}{\beta L (2\pi \anull)^2}  L \beta^2  e^{-2B} \;, \\
=&&- \frac{D}{ (2\pi \anull)^2} \int_0^\beta d\tau  \; \left[ e^{-2\widetilde{B}(\tau)}-e^{-2B}\right] -  \frac{\Sigma}{8}\;, 
\eea
where we have used the definition
\beq
\Sigma = u_c \sigma = u_c \frac{2D}{\pi^2\anull^2}e^{-2B}\;. 
\eeq
The integral in $f_3$ is problematic as it has power-law divergence. It can in principle be cut by the UV cutoff. Since we will neglect $I(\omega)$ everywhere, the consistent approach might be to neglect this integral altogether.\\

\noindent For $f_2$ we find:
\begin{align}
f_2 &= - \frac{1}{2 L} \sum_{q,\omega_n}  \frac{\hat{\Sigma}}{\frac{\omega_n^2}{v^2}+q^2+\hat{\Sigma}} + \frac{1}{L}\sum_{q>0} \frac{\hat{\sigma}}{q^2 + \hat{\Sigma}}  \nn \\
&= -\frac{v \hat{\Sigma}}{2L} \sum_{q > 0} \frac{ \coth (\beta v \sqrt{q^2+\hat{\Sigma}}/2)}{\sqrt{q^2+\hat{\Sigma}}} + + \frac{1}{L}\sum_{q>0} \frac{\hat{\Sigma}/u_c}{q^2 + \hat{\Sigma}} \;.
\end{align}

\noindent Finally, in order to compute $\Delta f_1$ we use the following formula for the determinant of a 1RSB matrix. If $A = (\tilde{a},a)$ is such a matrix, with break point $u_c$, then 
\beq
\frac{1}{n} \det A \sim \ln (\tilde{a}-a) + \frac{1}{u_c} \ln \frac{\tilde{a}-(1-u_c)a}{\tilde{a}-a}\quad \textrm{as} \quad n \to 0 \;. 
\eeq
Using standard formulas for the determinant of block matrices we find: 
\beq
\Delta f_1 = - \frac{2}{\beta L} \sum_{q >0} \ln \left[ \frac{\sinh(\beta  v q/2)}{\sinh(\beta  v \sqrt{q^2 + \hat{\Sigma}}/2)} \right] -\frac{1}{u_c \beta L} \sum_{q>0} \ln (1+\hat{\Sigma}/q^2)\;.
\eeq
In the limit $\beta \to \infty, L \to \infty$, we find
\beq
\Delta f =  \frac{v\hat{\Sigma}}{8\pi} -\frac{\sqrt{\hat{\Sigma}}}{2\beta u_c} +\frac{\sqrt{\hat{\Sigma}}}{4\beta u_c} -\frac{\hat{\Sigma}}{8}\frac{v}{\pi K}\;.
\eeq
At this point we can use the marginality condition, which yields,
\beq
\hat{\Sigma} = \frac{\pi K}{v} \frac{\sqrt{\hat{\Sigma}}}{u_c \beta}.
\eeq
Collecting all terms, we are left with
\beq
\Delta f = \frac{v\hat{\Sigma}}{8\pi}(1-3/K).
\eeq

\end{document}